\UseRawInputEncoding 
\documentclass[pra,twocolumn,superscriptaddress]{revtex4}
\usepackage{amsmath}
\usepackage{amssymb}
\usepackage{amsfonts}
\usepackage{graphicx}
\usepackage{textcomp}
\usepackage{lscape}
\usepackage{subcaption}
\usepackage{listings}
\usepackage{hhline}
\usepackage{multirow}
\usepackage{physics}
\usepackage{dcolumn}
\usepackage{xcolor}
\usepackage{bm}
\usepackage{appendix}
\usepackage{tikz}
\usepackage{caption}
\usetikzlibrary{quantikz2,backgrounds,fit,decorations.pathreplacing}
\providecommand{\U}[1]{\protect\rule{.1in}{.1in}}

\begin{document}
\preprint{APS/123-QED}

\title{A Measurement Device Independent Quantum Key Distribution protocol in the service of three users}

\author{Nikolaos Stefanakos}

\affiliation{Department of Informatics and Telecommunications, National and Kapodistrian
University of Athens, Panepistimiopolis, Ilisia, 15784, Greece}
\affiliation{Eulambia Advanced Technologies, Agiou Ioannou 24, Building Complex C, Ag. Paraskevi, 15342, Greece}

\author{Georgios Maragkopoulos}

\affiliation{Department of Informatics and Telecommunications, National and Kapodistrian
University of Athens, Panepistimiopolis, Ilisia, 15784, Greece}
\affiliation{Eulambia Advanced Technologies, Agiou Ioannou 24, Building Complex C, Ag. Paraskevi, 15342, Greece}

\author{ Aikaterini Mandilara}

\affiliation{Department of Informatics and Telecommunications, National and Kapodistrian
University of Athens, Panepistimiopolis, Ilisia, 15784, Greece}
\affiliation{Eulambia Advanced Technologies, Agiou Ioannou 24, Building Complex C, Ag. Paraskevi, 15342, Greece}

\author{ Dimitris Syvridis}

\affiliation{Department of Informatics and Telecommunications, National and Kapodistrian
University of Athens, Panepistimiopolis, Ilisia, 15784, Greece}
\affiliation{Eulambia Advanced Technologies, Agiou Ioannou 24, Building Complex C, Ag. Paraskevi, 15342, Greece}

\begin{abstract}
Quantum Key Distribution (QKD) is the only theoretically proven method for secure key distribution between two users. In this work, we propose and analyze a Measurement Device Independent (MDI) protocol designed to distribute keys among three users in a pairwise manner. Each user randomly selects a basis, encodes bit values in the phase of coherent states, and sends the resulting pulses to a central measurement unit (MU) composed of three beam splitters and three photon detectors. When the three pulses arrive simultaneously at the MU and under the condition of succesful detection of photons, a key bit is distributed to at least one pair of users. This protocol extends the foundational phase-encoding MDI protocol introduced by [K. Tamaki, et al., Phys. Rev. A 85, 042307 (2012)] to three users, but this comes at the cost of introducing a systematic error in the implementation of the honest protocol.

\end{abstract}
\maketitle
\section{Introduction}\label{S1}

Four decades after the introduction of the first Quantum Key Distribution (QKD) protocol, BB84 \cite{bib:bb84_protocol}, the field has evolved significantly, driven by the need for provable unconditional security in realistic use cases. The introduction of the first Measurement Device Independent (MDI) protocol \cite{bib:first_polarization_encoding_mdi_qkd_paper} marked a major advancement in the QKD field for two reasons. First, the measurement unit (MU) was moved to a third party, which can be untrusted and potentially controlled by an eavesdropper, but whose imperfections do not affect the security of the protocol. Second, the MDI protocol increased the achievable distance between two users.  This advancement is a key point for current cutting-edge protocols like Twin-Field \cite{bib:experimental_twin-field_quantum_key_distribution_through_sending_or_not_sending, bib:twin_field_quantum_key_distribution_over_830-km_fibre} and Mode-Pairing \cite{bib:mode_pairing_qkd}, which generalize the applicability of the first MDI protocols \cite{bib:first_polarization_encoding_mdi_qkd_paper, bib:first_phase_encoding__mdi_qkd_protocol}.

\begin{figure}[!b]
   \centering
    \includegraphics[width=1\linewidth]{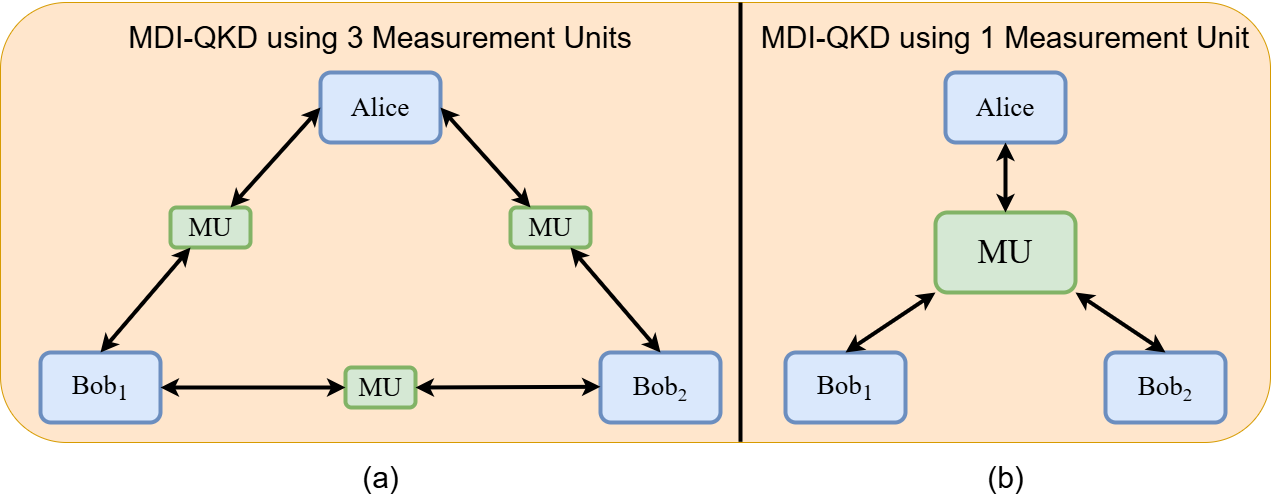}
    \captionsetup{justification=justified, font=small}
    \caption{\textit{(a)} A straightforward scheme for distributing keys to three users in a pairwise manner using MDI QKD. \textit{(b)} The scheme proposed in this work. Note that in scheme (b), the central MU is necessarily more complex than those in scheme (a). Assuming a fixed maximum average distance for undistorted quantum signal propagation, the maximum achievable distance between users in scheme (b) is reduced by a factor of $\sqrt{3}/2$ compared to scheme (a).}
    \label{fig:1}
\end{figure}

In this work, we extend the first of the two protocols proposed in the seminal paper "Phase Encoding Schemes for MDI QKD with Basis-Dependent Flaw" \cite{bib:first_phase_encoding__mdi_qkd_protocol} to a three-user scenario. Unlike protocols designed for Quantum Conference Key Agreement (QCKA) \cite{bib:review_2020_cka, bib:overcoming_fundamental_bounds_on_cka, bib:cka_in_quantum_network, bib:multi_user_mdi_ghz}—which aim to establish a shared key among all users—this protocol focuses on the pairwise distribution of keys among the three possible user pairs formed by Alice (A), Bob$_1$ (B$_1$), and Bob$_2$ (B$_2$). As in the original protocol, there is no need to distribute entangled states; instead, the users send their encoded coherent pulses to a central MU. When the MU announces a successful measurement, the three users reveal their encoding bases. The users with matching bases then append a bit to their shared key sequence. The key feature of the protocol is the use of a single MU instead of three, as would be required in a straightforward approach (see Fig.~\ref{fig:1}). An additional benefit is that base matching among all users is not necessary for pairwise key distribution, thereby reducing measurement discards by a factor of two. In the straightforward scenario, a user discards with a probability of $50\%$, while in the introduced scenario, the probability of discarding is reduced to $25\%$. However, this approach comes with practical trade-offs: it requires the simultaneous arrival of signals from all users at the MU, reduces the maximum achievable distance between users by a factor of $\sqrt{3}/2$ compared to the straightforward approach, and introduces a systematic error in the honest protocol, ultimately impacting the Secure Key Rate (SKR).

In this paper, we first present in Section~\ref{S2} the resources available to the users, the details of the MU unit, and the overall optical setup of the protocol. In the following Section (Section~\ref{S3}), we proceed with analytical derivations of the outcomes, considering two cases: a) \textit{Bases mismatch}, where one user encodes the bit value in a different basis than the other two, and b) \textit{Bases match}, where all users select the same encoding basis. The investigations aim to align the users' encodings with distinct measurement outcomes at the MU, ensuring that the encoded values remain private to the users, even when measurement outcomes and information about the bases are revealed. We also present the probability of success as a function of the intensity of the coherent pulses, along with the average probability of error for the proposed protocol.
Based on these results, we describe in Section~\ref{S4} the steps of a unified protocol for both bases mismatch and bases match scenarios. In Section~\ref{S5}, we provide preliminary information about the security of the protocol, drawing on elements from the works \cite{bib:first_phase_encoding__mdi_qkd_protocol} and \cite{bib:multiparty_qkd_protocol_without_entanglement}. Finally, in Section~\ref{S6}, we summarize the outcomes and discuss the perspectives of the proposed protocol.

\section{Resources and optical set-up}\label{S2}

\begin{figure}[!b]
    \centering
    \includegraphics[width=1\linewidth]{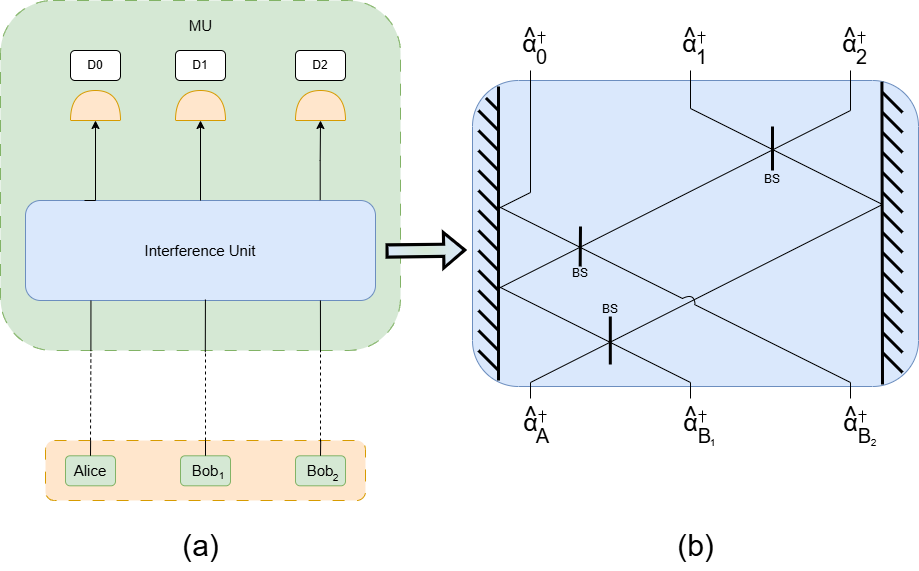}
    \captionsetup{justification=justified, font=small}
    \caption{The optical set-up  of the protocol: \textit{(a)} the overall setting,  \textit{(b)} the IU of the MU unit. \textit{BS} refers to  balanced beam splitter and $D_i$ to photon detector.}
    \label{fig:2}
\end{figure}

The goal is to design a three-user QKD protocol extending the two-user phase encoding scheme I from \cite{bib:first_phase_encoding__mdi_qkd_protocol}. We use the same encoding choices for the users as in \cite{bib:first_phase_encoding__mdi_qkd_protocol}. As in the two-user case, signals arriving at the MU undergo a unitary transformation, but now the output signals are measured by three photon detectors (see Fig.~\ref{fig:2}). The transformation induced by the Interference Unit (IU) is more complex than in \cite{bib:first_phase_encoding__mdi_qkd_protocol}, with the IU consisting of three balanced beam splitters. 

In more detail, each of the three users (A, B$_1$, and B$_2$) prepares and sends both a strong reference laser pulse and a weak coherent ``signal'' pulse. The reference pulse does not encode any information; it is used for polarization alignment of the three signals and for calculating the phase drift applied to the transmitted states due to fiber propagation. The signal pulse carries the encoded information and is described by $\vert \sqrt{\mu} e^{i\theta} \rangle$, where $\mu$ is the fixed mean photon number of the state throughout the protocol, and $\theta$ is the phase used for encoding.
Let the light modes of users A, B$_1$, and B$2$ be denoted as $\hat{a}_A^{\dagger}$, $\hat{a}_{B_1}^{\dagger}$, and $\hat{a}_{B_2}^{\dagger}$, respectively. Each user randomly chooses a bit value and the encoding basis. For the $X$ basis, a bit value of $0$ ($1$) is encoded with phase $0$ ($\pi$), while for the $Y$ basis, a bit value of $0$ ($1$) is encoded with phase $\pi/2$ ($3\pi/2$). These two bases are not equivalent in the phase-encoding scheme, as $\rho_X = \frac{1}{2} \vert \sqrt{\mu} \rangle \langle \sqrt{\mu} \vert + \frac{1}{2} \vert -\sqrt{\mu} \rangle \langle -\sqrt{\mu} \vert$ is distinguishable from $\rho_Y = \frac{1}{2} \vert i \sqrt{\mu} \rangle \langle i \sqrt{\mu} \vert + \frac{1}{2} \vert -i\sqrt{\mu} \rangle \langle -i\sqrt{\mu} \vert$. This basis-dependent flaw, which could potentially be exploited by an eavesdropper, can be quantified using a simple measure of fidelity between density matrices with the same degree of mixedness:
$\mathrm{Tr}\left(\rho_X \rho_Y \right)/\mathrm{Tr}\left(\rho_X^2 \right)=1/cosh\left(2\mu\right)$.
For a mean photon number $\mu < 0.3$, the fidelity of the two density matrices remains above 0.84, and in this work, we assume  low-amplitude coherent states for encoding the bit values.

Following the flow of  Fig.~\ref{fig:2}~(a), the pulses sent by the three users, propagate at equal fiber lengths   to  arrive simultaneously at the MU. The MU  is composed by  an IU which applies a rotation to the input modes, $\hat{a}_A^{\dagger}$, $\hat{a}_{B_1}^{\dagger}$, $\hat{a}_{B_2}^{\dagger}$, and outputs the modes $\hat{a}_0^{\dagger}$, $\hat{a}_{1}^{\dagger}$, $\hat{a}_{2}^{\dagger}$. The states on the latter modes are then guided to the photon detectors $D0$, $D1$ and $D2$ accordingly.  To build the protocol we take the usual assumption that a photon detector has two states: ``fire'', detecting the   presence of at least one photon in the respected output mode and ``not fire''.

The unitary operation on the input modes corresponding to the IU in Fig.~\ref{fig:2}~(b) can be described as a rotation $\hat{R}$ applied to the input modes by the IU:
\begin{equation}
\mathbf{R} = e^{\phi_x \hat{L}_x} \cdot e^{\phi_y \hat{L}_y} \cdot e^{\phi_z \hat{L}_z} \label{eq_test}
\end{equation}
where $\{\hat{L}_x, \hat{L}_y, \hat{L}_z\}$ are the $3\times 3 $ generators of orthogonal group $O(3)$ and $\phi_x=\phi_y=\phi_z=\pi/4$. The structure of the IU naturally extends the IU from \cite{bib:first_phase_encoding__mdi_qkd_protocol}, selected among other possible configurations, as it leads to measurement outcomes that meet the basic requirements of the protocol.

\section{Relating users' input to measurement outcomes}\label{S3}

After defining the possible states sent by the users and the optical setup in Fig.\ref{fig:2}, we calculate the states exiting the IU and reaching the detectors. The detection outcomes are not mutually exclusive, so we identify detection types that maximize the probability of correct detection while minimizing misdetections. Once the MU operator announces the detection type, the users publicly reveal their encoding bases, leading to two scenarios: a) \textit{Bases Mismatch}, where one user encodes in a different basis, and b) \textit{Bases Match}, where all users encode in the same basis. We analyze these two cases separately. In half of the cases, the bases announcement requires some users to flip their encoded bit, which is discussed here, even though the protocol steps are presented more clearly in Section\ref{S4}.

\subsection{Bases mismatch}

\textit{Bases mismatch} describes any of the following six basis choices of the users A, B$_1$ and B$_2$: $\left\{ \textrm{XXY, YYX, XYX, YXY, YXX, XYY} \right\}$. Furthermore one can pair the triplets which coincide if X $\leftrightarrow$ Y, e.g., $\left\{ \textrm{XXY, YYX} \right\}$, since both options output coherent states of the same amplitude.
For the first pair, assuming that the amplitude of the input states is $\sqrt{\mu}$, in Table~\ref{tab:1} we provide the amplitudes of the coherent states reaching the detectors for all possible choices of encoding. The Table~\ref{tab:1} illustrates that each output state provides a probability for each detector to either fire or remain inactive. Thus we are obliged to `enforce' to each scenario a detection type keeping though in mind not only that a detection might not occur but more importantly that a misdetection can happen as well. For instance, an input state characterized on Table~\ref{tab:1} as Type 0 can lead with some probability to detection Type 1 and vice versa.

\textit{Bases mismatch} refers to any of the following six possible basis choices for users A, B$_1$, and B$_2$: $\left\{ \textrm{XXY, YYX, XYX, YXY, YXX, XYY} \right\}$. These can be paired by swapping X $\leftrightarrow$ Y (e.g., $\left\{ \textrm{XXY, YYX} \right\}$), as both options yield coherent states with the same amplitude. For the first pair, assuming the amplitude of the input states is $\sqrt{\mu}$, Table~\ref{tab:1} shows the amplitudes of the coherent states reaching the detectors for all encoding choices. The table illustrates that each output state results in a probability for each detector to either fire or remain inactive. We must therefore `enforce' a detection type for each scenario, keeping in mind not only that a detection might not occur, but also that misdetections can happen (contrary to \cite{bib:first_phase_encoding__mdi_qkd_protocol}).  In Appendix we provide the tables for the rest of the triplets.

\begin{figure}[b]
\centering
    \includegraphics[width=0.8\linewidth]{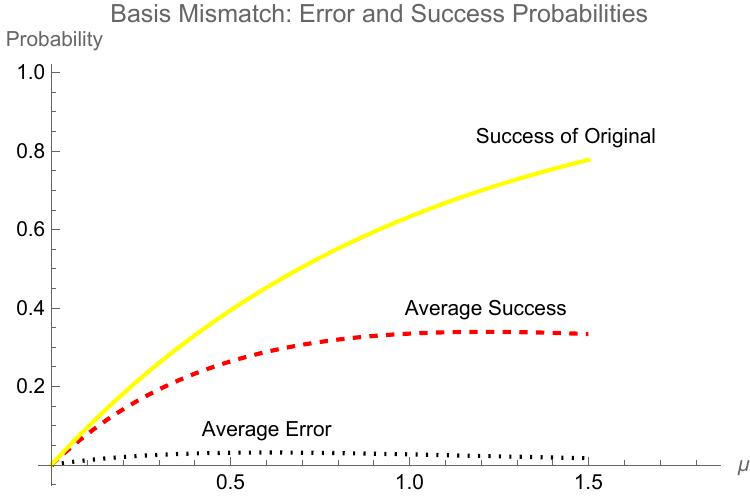}
    \captionsetup{justification=justified, font=small}
    \caption{Bases mismatch between the three users, where a bit is distributed to the pair of users with matching bases.
    Dashed red line: Average probability of a successful detection event, assuming perfect detectors.
    Dotted black line: Average probability of a wrong detection type, leading to a bit error in the pairwise distributed key.
    Solid yellow line: Probability of successful detection for the original protocol in \cite{bib:first_phase_encoding__mdi_qkd_protocol}.
    Horizontal axis: Intensity $\mu$ of the coherent states prepared by the users.}
    \label{fig:3}
\end{figure}

In Fig.~\ref{fig:3}  we present the average success probability of correct detection over all six bases triplets and phase encodings, assuming perfect detectors. The average  error is presented in the same graph. This concerns   an honest implementation of the protocol, and for this reason we  refer to  it as \textit{ systematic error} to differentiate it from errors due to eavesdropping or imperfection on devices/links.   

\begin{table*}
\captionsetup{justification=justified, font=small, width=0.97\textwidth}
\caption{Bases mismatch: XXY and YYX  bases choices of the users A, B$_1$ and B$_2$. For each possible encoding on a bases triplet, the \textit{amplitudes} of the coherent states  reaching the detectors are listed (rounded to the second decimal digit). We attribute two different types of detection: Type 0 when the matching pair of users encodes the same bit value (phase) and 1  when the users send encoded pulses with phase difference of $\pi$. In the latter case, one of the user needs to flip her/his  registered bit value to create a common bit in the shared key. The symbol  $\wedge ~~\wedge$ on the same row signifies  simultaneous clicks on  detectors (D$_1$ and D$_2$).}
\label{tab:1}
\begin{subtable}{1\textwidth}
    \centering
    \setlength{\arrayrulewidth}{0.25mm}
    \setlength{\tabcolsep}{7pt}
    \renewcommand{\arraystretch}{1.6}
    \begin{tabular}{|ccc||ccc||ccc||c||c|}
        \hline
        \multicolumn{3}{|c||}{\textbf{Users}} & \multicolumn{3}{c||}{\textbf{IU output states}} & \multicolumn{3}{c||}{\textbf{Detector click}} & \multicolumn{1}{c||}{\textbf{Detection}} & \multicolumn{1}{c|}{\textbf{Required}} \\ \hhline{|-|-|-|-|-|-|-|-|-|}
        
       \multicolumn{1}{|c|}{$\textbf{A:~X}$} & \multicolumn{1}{c|}{$\textbf{B}_\textbf{1}\textbf{:~X}$} & $\textbf{B}_\textbf{2}\textbf{:~Y}$ & \multicolumn{1}{c|}{$\hat{a}_0^{\dagger}$} & \multicolumn{1}{c|}{$\hat{a}_1^{\dagger}$} & \multicolumn{1}{c||}{$\hat{a}_2^{\dagger}$} & \multicolumn{1}{c|}{\textbf{D0}} & \multicolumn{1}{c|}{\textbf{D1}} & \multicolumn{1}{c||}{\textbf{D2}} & \multicolumn{1}{c||}{\textbf{Type}} & \multicolumn{1}{c|}{\textbf{Actions}} \\ \hhline{|=|=|=#=|=|=#=|=|=#=#=|}

        \multicolumn{1}{|c|}{0} & \multicolumn{1}{c|}{0} & $\frac{\pi}{2}$ or $\frac{3\pi}{2}$ & \multicolumn{1}{c|}{\multirow{2}{*}{$0.71\sqrt{\mu}$}} & \multicolumn{1}{c|}{\multirow{2}{*}{$1.12\sqrt{\mu}$}} & \multicolumn{1}{c||}{\multirow{2}{*}{$1.12\sqrt{\mu}$}} & \multicolumn{1}{c|}{\multirow{2}{*}{}} & \multicolumn{1}{c|}{\multirow{2}{*}{$\wedge$}} & \multicolumn{1}{c||}{\multirow{2}{*}{$\wedge$}} & \multicolumn{1}{c||}{\multirow{2}{*}{Type 0}} & \multicolumn{1}{c|}{\multirow{2}{*}{-}}\\ \hhline{|-|-|-|}
        
        \multicolumn{1}{|c|}{$\pi$} & \multicolumn{1}{c|}{$\pi$} & $\frac{\pi}{2}$ or $\frac{3\pi}{2}$ & \multicolumn{1}{c|}{} & \multicolumn{1}{c|}{} & \multicolumn{1}{c||}{} & \multicolumn{1}{c|}{} & \multicolumn{1}{c|}{} & \multicolumn{1}{c||}{} & \multicolumn{1}{c||}{} & \multicolumn{1}{c|}{}\\ \hhline{|=|=|=#=|=|=#=|=|=#=#=|}
        
        \multicolumn{1}{|c|}{0} & \multicolumn{1}{c|}{$\pi$} & $\frac{\pi}{2}$ or $\frac{3\pi}{2}$ & \multicolumn{1}{c|}{\multirow{2}{*}{$1.22\sqrt{\mu}$}} & \multicolumn{1}{c|}{\multirow{2}{*}{$0.87\sqrt{\mu}$}} & \multicolumn{1}{c||}{\multirow{2}{*}{$0.87\sqrt{\mu}$}} & \multicolumn{1}{c|}{\multirow{2}{*}{$\wedge$}} & \multicolumn{1}{c|}{\multirow{2}{*}{}} & \multicolumn{1}{c||}{\multirow{2}{*}{}} & \multicolumn{1}{c||}{\multirow{2}{*}{Type 1}} & \multicolumn{1}{c|}{\multirow{2}{*}{$\textbf{B}_\textbf{1}$ flips}}\\ \hhline{|-|-|-|}
        
        \multicolumn{1}{|c|}{$\pi$} & \multicolumn{1}{c|}{0} & $\frac{\pi}{2}$ or $\frac{3\pi}{2}$ & \multicolumn{1}{c|}{} & \multicolumn{1}{c|}{} & \multicolumn{1}{c||}{} & \multicolumn{1}{c|}{} & \multicolumn{1}{c|}{} & \multicolumn{1}{c||}{} & \multicolumn{1}{c||}{} & \multicolumn{1}{c|}{}\\ \hhline{|-|-|-|-|-|-|-|-|-|-|-|}
    \end{tabular}
\end{subtable}

\bigskip
\begin{subtable}{1\textwidth}
    \centering
    \setlength{\arrayrulewidth}{0.25mm}
    \setlength{\tabcolsep}{7.1pt}
    \renewcommand{\arraystretch}{1.6}
    \begin{tabular}{|ccc||ccc||ccc||c||c|}
        \hline
        \multicolumn{3}{|c||}{\textbf{Users}} & \multicolumn{3}{c||}{\textbf{IU output states}} & \multicolumn{3}{c||}{\textbf{Detector click}} & \multicolumn{1}{c||}{\textbf{Detection}} & \multicolumn{1}{c|}{\textbf{Required}} \\ \hhline{|-|-|-|-|-|-|-|-|-|}
        
        \multicolumn{1}{|c|}{$\textbf{A:~Y}$} & \multicolumn{1}{c|}{$\textbf{B}_\textbf{1}\textbf{:~Y}$} & $\textbf{B}_\textbf{2}\textbf{:~X}$  & \multicolumn{1}{c|}{$\hat{a}_0^{\dagger}$} & \multicolumn{1}{c|}{$\hat{a}_1^{\dagger}$} & \multicolumn{1}{c||}{$\hat{a}_2^{\dagger}$} & \multicolumn{1}{c|}{\textbf{D0}} & \multicolumn{1}{c|}{\textbf{D1}} & \multicolumn{1}{c||}{\textbf{D2}} & \multicolumn{1}{c||}{\textbf{Type}} & \multicolumn{1}{c|}{\textbf{Actions}} \\ \hhline{|=|=|=#=|=|=#=|=|=#=#=|}

        \multicolumn{1}{|c|}{$\frac{\pi}{2}$} & \multicolumn{1}{c|}{$\frac{\pi}{2}$} & 0 or $\pi$ & \multicolumn{1}{c|}{\multirow{2}{*}{$0.71\sqrt{\mu}$}} & \multicolumn{1}{c|}{\multirow{2}{*}{$1.12\sqrt{\mu}$}} & \multicolumn{1}{c||}{\multirow{2}{*}{$1.12\sqrt{\mu}$}} & \multicolumn{1}{c|}{\multirow{2}{*}{}} & \multicolumn{1}{c|}{\multirow{2}{*}{$\wedge$}} & \multicolumn{1}{c||}{\multirow{2}{*}{$\wedge$}} & \multicolumn{1}{c||}{\multirow{2}{*}{Type 0}} & \multicolumn{1}{c|}{\multirow{2}{*}{-}}\\ \hhline{|-|-|-|}
        
        \multicolumn{1}{|c|}{$\frac{3\pi}{2}$} & \multicolumn{1}{c|}{$\frac{3\pi}{2}$} & 0 or $\pi$ & \multicolumn{1}{c|}{} & \multicolumn{1}{c|}{} & \multicolumn{1}{c||}{} & \multicolumn{1}{c|}{} & \multicolumn{1}{c|}{} & \multicolumn{1}{c||}{} & \multicolumn{1}{c||}{} & \multicolumn{1}{c|}{}\\ \hhline{|=|=|=#=|=|=#=|=|=#=#=|}
        
        \multicolumn{1}{|c|}{$\frac{\pi}{2}$} & \multicolumn{1}{c|}{$\frac{3\pi}{2}$} & 0 or $\pi$ & \multicolumn{1}{c|}{\multirow{2}{*}{$1.22\sqrt{\mu}$}} & \multicolumn{1}{c|}{\multirow{2}{*}{$0.87\sqrt{\mu}$}} & \multicolumn{1}{c||}{\multirow{2}{*}{$0.87\sqrt{\mu}$}} & \multicolumn{1}{c|}{\multirow{2}{*}{$\wedge$}} & \multicolumn{1}{c|}{\multirow{2}{*}{}} & \multicolumn{1}{c||}{\multirow{2}{*}{}} & \multicolumn{1}{c||}{\multirow{2}{*}{Type 1}} & \multicolumn{1}{c|}{\multirow{2}{*}{$\textbf{B}_\textbf{1}$ flips}}\\ \hhline{|-|-|-|}
        
        \multicolumn{1}{|c|}{$\frac{3\pi}{2}$} & \multicolumn{1}{c|}{$\frac{\pi}{2}$} & 0 or $\pi$ & \multicolumn{1}{c|}{} & \multicolumn{1}{c|}{} & \multicolumn{1}{c||}{} & \multicolumn{1}{c|}{} & \multicolumn{1}{c|}{} & \multicolumn{1}{c||}{} & \multicolumn{1}{c||}{} & \multicolumn{1}{c|}{}\\ \hhline{|-|-|-|-|-|-|-|-|-|-|-|}
    \end{tabular}
\end{subtable}
\end{table*}

\subsection{Bases match}
In the event that all users encode the information on the same basis, e.g., XXX, we identify four different  patterns of detection, presented in Table~\ref{tab:2}. As for Bases mismatch, in  Table~\ref{tab:2} we relate the inputs to the outputs, the detection types and the required actions so that each pair adds a bit on its pairwise key -- bit-string.

\begin{figure}[!b]
    \centering
    \includegraphics[width=0.8\linewidth]{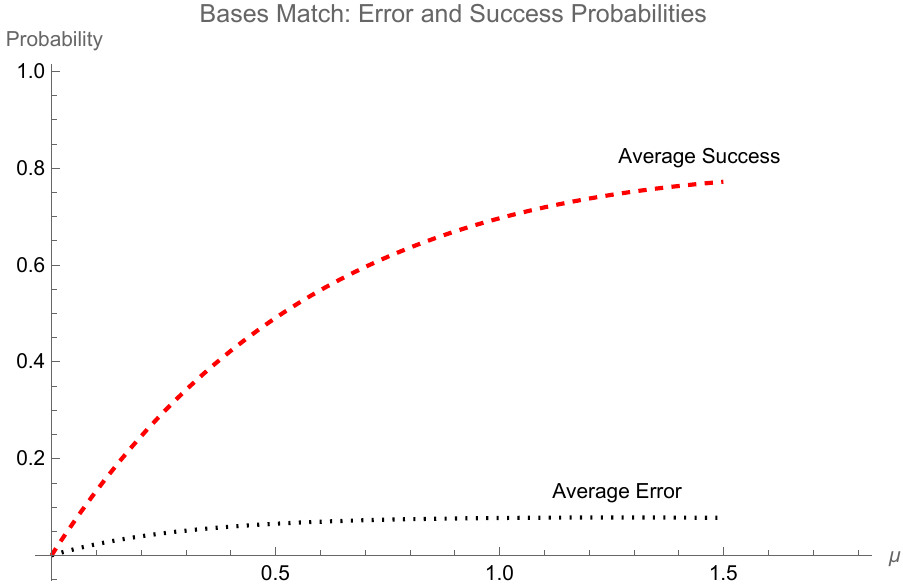}
    \captionsetup{justification=justified, font=small}
    \caption{Bases match. Using the information in Table~\ref{tab:2}, we calculate the average probability of a successful detection for Types 0 or 1, as well as the probability of a systematic error for these detection types. For the latter, we average the probabilities that an input of Type 1 (0), 3, or 4 in Table~\ref{tab:2} results in a detection of Type 0 (1).
}
    \label{fig:4}
\end{figure}

We calculate the probability of systematic errors for an honest implementation, as shown in Table~\ref{tab:2}. In calculations not presented here, we observe a significant systematic error of about $20\%$ for $\mu \approx 0.4$, primarily due to the overlap between detection outcomes of types 3 and 4 with type 0. This high systematic error leads to a Bit Error Rate (BER) of approximately $40\%$, rendering the \textit{Bases match} case of the protocol impractical. To address this, we exclude detection types 3 and 4, as doing so, significantly reduces the systematic error without compromising the probability of a successful implementation, making the BER more tolerable. In Fig.~\ref{fig:4}, we plot the probability of a successful detection for Types 0 and 1, along with the corresponding systematic error introduced by other detection types. Finally, Table~\ref{tab:3} summarizes the admissible detection types for each basis triplet.

It is important to note that a Bases Match (excluding detection types 3 and 4) occurs with a probability of $1/8$ across all cases. This probability is further adjusted by the probability of successful implementation, as shown in Fig.~\ref{fig:4}. The low overall probability of success, coupled with the additional systematic error, makes the current protocol unsuitable for implementing QCKA. For a more effective approach, we refer interested readers to more sophisticated protocols, such as the one in \cite{bib:overcoming_fundamental_bounds_on_cka}.

\begin{table*}
    \captionsetup{justification=justified, font=small,width=0.95\textwidth}
    \caption{Bases match: XXX bases choices for the users. For each possible
    users' input we list the amplitudes of the coherent states reaching the detectors and identify
    admissible types of detection.  The results are identical for the triplet YYY.}
    \label{tab:2}
    \centering
    \setlength{\arrayrulewidth}{0.25mm}
    \setlength{\tabcolsep}{8pt}
    \renewcommand{\arraystretch}{1.25}
    \begin{tabular}{|ccc||ccc||ccc||c||c|}
        \hline
        \multicolumn{3}{|c||}{\textbf{Users}} & \multicolumn{3}{c||}{\textbf{IU output states}} & \multicolumn{3}{c||}{\textbf{Detector click}} & \multicolumn{1}{c||}{\textbf{Detection}} & \multicolumn{1}{c|}{\textbf{Required}} \\ \hhline{|-|-|-|-|-|-|-|-|-|}
        \multicolumn{1}{|c|}{ $\textbf{A}$ } & \multicolumn{1}{c|}{$\textbf{B}_\textbf{1}$} & $\textbf{B}_\textbf{2}$ & \multicolumn{1}{c|}{$\hat{a}_0^{\dagger}$} & \multicolumn{1}{c|}{$\hat{a}_1^{\dagger}$} & \multicolumn{1}{c||}{$\hat{a}_2^{\dagger}$} & \multicolumn{1}{c|}{\textbf{D0}} & \multicolumn{1}{c|}{\textbf{D1}} & \multicolumn{1}{c||}{\textbf{D2}} & \multicolumn{1}{c||}{\textbf{Type}} & \multicolumn{1}{c|}{\textbf{Actions}} \\ \hhline{|=|=|=#=|=|=#=|=|=#=#=|}
        
        \multicolumn{1}{|c|}{0} & \multicolumn{1}{c|}{$\pi$} & $\pi$ & \multicolumn{1}{c|}{\multirow{2}{*}{$0.29\sqrt{\mu}$}} & \multicolumn{1}{c|}{\multirow{2}{*}{$1.20\sqrt{\mu}$}} & \multicolumn{1}{c||}{\multirow{2}{*}{$1.20\sqrt{\mu}$}} & \multicolumn{1}{c|}{\multirow{2}{*}{}} & \multicolumn{1}{c|}{\multirow{2}{*}{$\wedge$}} & \multicolumn{1}{c||}{\multirow{2}{*}{$\wedge$}} & \multicolumn{1}{c||}{\multirow{2}{*}{Type 0}} & \multicolumn{1}{c|}{\multirow{2}{*}{$\textbf{B}_\textbf{1}$,$\textbf{B}_\textbf{2}$ flip}} \\ \hhline{|-|-|-|}
        \multicolumn{1}{|c|}{$\pi$} & \multicolumn{1}{c|}{0} & 0 & \multicolumn{1}{c|}{} & \multicolumn{1}{c|}{} & \multicolumn{1}{c||}{} & \multicolumn{1}{c|}{} & \multicolumn{1}{c|}{} & \multicolumn{1}{c||}{} & \multicolumn{1}{c||}{} & \multicolumn{1}{c|}{} \\ \hhline{|=|=|=#=|=|=#=|=|=#=#=|}

        \multicolumn{1}{|c|}{0} & \multicolumn{1}{c|}{$\pi$} & 0 & \multicolumn{1}{c|}{\multirow{2}{*}{$1.70\sqrt{\mu}$}} & \multicolumn{1}{c|}{\multirow{2}{*}{$0.22\sqrt{\mu}$}} & \multicolumn{1}{c||}{\multirow{2}{*}{$0.22\sqrt{\mu}$}} & \multicolumn{1}{c|}{\multirow{2}{*}{$\wedge$}} & \multicolumn{1}{c|}{\multirow{2}{*}{}} & \multicolumn{1}{c||}{\multirow{2}{*}{}} & \multicolumn{1}{c||}{\multirow{2}{*}{Type 1}} & \multicolumn{1}{c|}{\multirow{2}{*}{$\textbf{B}_\textbf{1}$ flips}} \\ \hhline{|-|-|-|}
        \multicolumn{1}{|c|}{$\pi$} & \multicolumn{1}{c|}{0} & $\pi$ & \multicolumn{1}{c|}{} & \multicolumn{1}{c|}{} & \multicolumn{1}{c||}{} & \multicolumn{1}{c|}{} & \multicolumn{1}{c|}{} & \multicolumn{1}{c||}{} & \multicolumn{1}{c||}{} & \multicolumn{1}{c|}{} \\ \hhline{|=|=|=#=|=|=#=|=|=#=#=|}

        \multicolumn{1}{|c|}{0} & \multicolumn{1}{c|}{0} & $\pi$ & \multicolumn{1}{c|}{\multirow{2}{*}{$0.71\sqrt{\mu}$}} & \multicolumn{1}{c|}{\multirow{2}{*}{$1.5\sqrt{\mu}$}} & \multicolumn{1}{c||}{\multirow{2}{*}{$0.5\sqrt{\mu}$}} & \multicolumn{1}{c|}{\multirow{2}{*}{}} & \multicolumn{1}{c|}{\multirow{2}{*}{$\wedge$}} & \multicolumn{1}{c||}{\multirow{2}{*}{}} & \multicolumn{1}{c||}{\multirow{2}{*}{Type 3}} & \multicolumn{1}{c|}{\multirow{2}{*}{$\textbf{B}_\textbf{2}$ flips}} \\ \hhline{|-|-|-|}
        \multicolumn{1}{|c|}{$\pi$} & \multicolumn{1}{c|}{$\pi$} & 0 & \multicolumn{1}{c|}{} & \multicolumn{1}{c|}{} & \multicolumn{1}{c||}{} & \multicolumn{1}{c|}{} & \multicolumn{1}{c|}{} & \multicolumn{1}{c||}{} & \multicolumn{1}{c||}{} & \multicolumn{1}{c|}{} \\ \hhline{|=|=|=#=|=|=#=|=|=#=#=|}
        
        \multicolumn{1}{|c|}{0} & \multicolumn{1}{c|}{0} & 0 & \multicolumn{1}{c|}{\multirow{2}{*}{$0.71\sqrt{\mu}$}} & \multicolumn{1}{c|}{\multirow{2}{*}{$0.5\sqrt{\mu}$}} & \multicolumn{1}{c||}{\multirow{2}{*}{$1.5\sqrt{\mu}$}} & \multicolumn{1}{c|}{\multirow{2}{*}{}} & \multicolumn{1}{c|}{\multirow{2}{*}{}} & \multicolumn{1}{c||}{\multirow{2}{*}{$\wedge$}} & \multicolumn{1}{c||}{\multirow{2}{*}{Type 4}} & \multicolumn{1}{c|}{\multirow{2}{*}{-}} \\ \hhline{|-|-|-|}
        \multicolumn{1}{|c|}{$\pi$} & \multicolumn{1}{c|}{$\pi$}& $\pi$ & \multicolumn{1}{c|}{} & \multicolumn{1}{c|}{} & \multicolumn{1}{c||}{} & \multicolumn{1}{c|}{} & \multicolumn{1}{c|}{} & \multicolumn{1}{c||}{} & \multicolumn{1}{c||}{} & \multicolumn{1}{c|}{} \\ \hhline{|-|-|-|-|-|-|-|-|-|-|-|}
    \end{tabular}
\end{table*}

\begin{table}[b]
    \centering
    \captionsetup{justification=justified, font=small}
    \caption{Summary of admissible detection types --bases mismatch and match. After announcing the detection type and selected bases, an event is discarded if the triplet of bases does not match the detection type as shown in this table.
    The occurrence of two  $\wedge$ along a row implies that \textit{both} detectors  simultaneously fire.}
    \label{tab:3}
    \begin{tabular}{|c||c||c|c|c|}
    \hline
      Detection Type& Bases Triplet (A, B$_1$, B$_2$) & D0  & D1 & D2  \\ \hline \hline
        0 &XXY, YYX, XXX, YYY& & $\wedge$&$\wedge$  \\ \hline
        1 &XXY, YYX, XXX, YYY&$\wedge$ & &   \\  \hline 
       2 & XXY, YYX & $\wedge$ & &$\wedge$  \\ \hline
       3 & XXY, YYX & & $\wedge$ &     \\  \hline 
        4 & XYX, YXY& & & $\wedge$ \\ \hline
        5 &XYX, YXY & $\wedge$ & $\wedge$&    \\ \hline \hline    
    \end{tabular}
\end{table}

\section{The steps of the protocol}\label{S4}

The analysis in the previous section prepares us to present the steps of the protocol for distributing keys. In the case of a Bases mismatch, the protocol distributes keys to a single pair of users. In the case of a Bases match, keys are distributed to all three pairs, with only Types 0 and 1 of detection being considered.
\begin{enumerate}
 \item Each user randomly selects a bit value ($b := 0,~1$) and a basis ($B := X,~Y$). They create and send a reference pulse followed by a coherent pulse with intensity $\mu$, whose phase $\phi$ is modulated according to the chosen bit value and basis as follows: \begin{itemize} \item $(b = 0, B = X) \rightarrow \phi = 0$ \item $(b = 1, B = X) \rightarrow \phi = \pi$ \item $(b = 0, B = Y) \rightarrow \phi = \pi/2$ \item $(b = 1, B = Y) \rightarrow \phi = 3\pi/2$ \end{itemize}
    
\item The signal pulses propagate through the fibers and arrive simultaneously at the MU, where they pass through the IU and are measured by the detectors (see Fig.~\ref{fig:2}). The measurement outcome is considered successful if: \textit{a)} one detector fires, or \textit{b)} two detectors fire simultaneously (see Table~\ref{tab:3}). If the outcome is successful, the detection type is also announced. If the measurement is unsuccessful, the users discard their data and restart from Step 1.
    
  \item Each user announces the basis used for encoding. If the triplet of bases does not match the detection type in Table~\ref{tab:3}, they discard their data and restart the process. If the bases match, the pair(s) with the matching bases generate a bit for their shared key by following the actions outlined in Tables~\ref{tab:1}-\ref{tab:2}, \ref{tab:a_1}-\ref{tab:a_2}.    
\end{enumerate}

\begin{figure}[!b]
    \centering
    \includegraphics[width=0.8\linewidth]{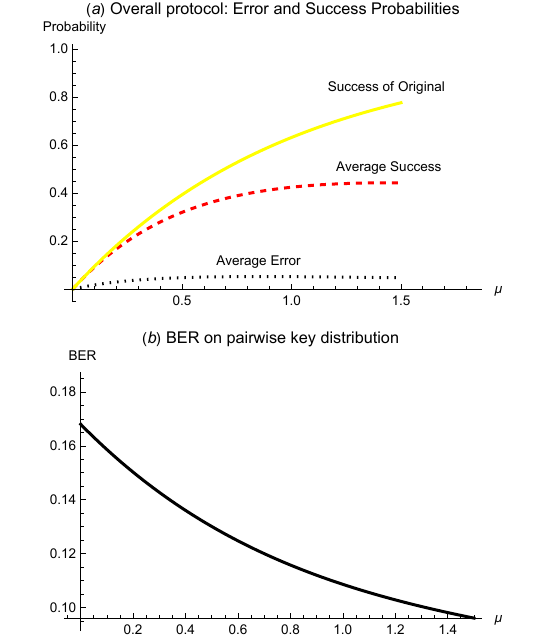}
    \captionsetup{justification=justified, font=small}
    \caption{\textit{(a)} Average probability of success (dashed line) and systematic error (dotted line) for the overall protocol described in Section~\ref{S4} versus the intensity $\mu$ of the coherent states prepared by the users. The probability of successful detection for the protocol in \cite{bib:first_phase_encoding__mdi_qkd_protocol} is also shown (solid line). \textit{(b)} BER for an honest implementation of the protocol. All lines in the figure refer to key distribution between a single pair of users. }
    \label{fig:5}
\end{figure}

 The users repeat the steps of the protocol until they generate pairwise keys of sufficient length for their needs. They then proceed to estimate two important parameters of the channel: the BER and the Phase Error Rate. The average probabilities for a successful detection and for systematic error in the protocol are presented in Fig.~\ref{fig:5}.

\section{Is the  protocol secure?}\label{S5}
We offer only a partial answer to this question. By decomposing the protocol into the Bases Mismatch and Bases Match cases, the security of the first part directly follows from the security proofs in \cite{bib:first_phase_encoding__mdi_qkd_protocol}. For the Bases Match case, however, we can only establish a non-rigorous connection to concepts derived in \cite{bib:multiparty_qkd_protocol_without_entanglement}.

\subsection{Bases Mismatch}
Assuming the protocol applies only in the Bases Mismatch case, the plots in Fig.~\ref{fig:3} represent the probability of successfully distributing a key to a pair of users and the associated systematic error. From these plots, it is clear that the key generation rate for the Mismatch protocol is lower than that of the one in \cite{bib:first_phase_encoding__mdi_qkd_protocol}, and the presence of a systematic error further reduces the SKR. However, the security parameters and the formula for estimating the SKR from \cite{bib:first_phase_encoding__mdi_qkd_protocol} still apply here, with the necessary adjustment for the inclusion of the systematic error factor in the estimations.

In more detail, for the Bases Mismatch protocol, the states sent by the three users are unentangled. As seen in Tables~\ref{tab:1}, \ref{tab:a_1}, and \ref{tab:a_2}, the bit value of the unmatched user does not correlate with the key of the matched users. Therefore, if the unmatched state is attributed to Eve, it does not increase her knowledge or influence over the protocol. One can re-design the virtual protocol from Section~5 of \cite{bib:first_phase_encoding__mdi_qkd_protocol} by taking the density matrix representing the ensemble of two different encodings for the unmatched user and putting it in product with the states of the paired users and Eve. This approach preserves $\Delta_{ini}$, the key security parameter of the protocol that quantifies the basis mismatch flaw.

 On the other hand, the input of the unmatched state introduces noise into the output of the paired users, thereby inducing the aforementioned systematic error (Fig.~\ref{fig:3}). To calculate the secure key generation rate, one can still apply the formula (10) from \cite{bib:first_phase_encoding__mdi_qkd_protocol}, with the bit error rates, $\delta_X$ and $\delta'{_Y}$, now augmented due to the systematic error, and the success probability $\gamma_{suc}$ reduced. Finally, error correction and privacy amplification can be applied independently to each pair of users/keys, just as in a typical QKD protocol.

\subsection{Bases Match}

 In the combined protocol of Section~\ref{S4}, the Bases Match scenario (excluding detection Types 3 and 4) contributes to $1/4$ of the key sequence for each pair of users. When treated as an independent protocol, the probabilities of successful detection and erroneous outcomes are shown in Fig.~\ref{fig:4}. The security proof for the Bases Match protocol would require an extension of the proofs in \cite{bib:first_phase_encoding__mdi_qkd_protocol} to accommodate three users. This extension is quite complex, as it would require treating the protocol as a QCKA protocol \cite{bib:Multi-partite}.

However, we have deliberately structured the protocol (see Tables~\ref{tab:1}-\ref{tab:2}, \ref{tab:a_1}-\ref{tab:a_2}) so that systematic errors only affect the bit strings of users $B_1$ and $B_2$, with user $A$ serving as the reference. If we focus solely on the systematic error and exclude other sources of noise or eavesdropping on Alice's signal, the error correction and hashing procedures from \cite{bib:multiparty_qkd_protocol_without_entanglement} can be applied to mitigate the impact of this systematic error. Under this assumption—that Alice's signal experiences no noise or eavesdropping—the formula in \cite{bib:multiparty_qkd_protocol_without_entanglement} can also be used to estimate the SKR for this part of the protocol.

\section{Discussion}\label{S6}

In this work, we developed a QKD protocol designed to serve three users in a pairwise manner. The protocol is built upon the MDI framework, offering several advantages, including its centralized configuration with a central MU and a star topology for the users. A key benefit of our protocol is that it significantly reduces the discard rate caused by bases mismatch—by almost a factor of two—when compared to a straightforward approach involving three separate MUs.

However, several open questions remain. The completion of the security proof for the protocol is still pending, and further investigation is needed to assess its resilience against noise, detector imperfections, and other practical limitations. Additionally, it would be valuable to explore whether this protocol can be simplified or extended to accommodate more users, or if the single-MU design presents any inherent bottlenecks. These aspects provide important directions for future research in the development of scalable and secure QKD protocols.

\begin{appendix}
\section{Bases Mismatch: tables for (YXY, XYX) and (XYY, YXX) bases triplets}

In the main text we provide the Tables \ref{tab:1}-\ref{tab:2}, for XXY, YYX, XXX and YYY bases triplets. In Tables~\ref{tab:a_1}-\ref{tab:a_2} we provide the information for the rest of the triplets. The calculations have been performed using basic elements of quantum optics \cite{bib:measuring_the_quantum_state_of_light}.

\begin{table*}
\caption{Bases mismatch: YXY and XYX  bases choices of the users A, B$_1$ and B$_2$. }
\label{tab:a_1}
\begin{subtable}{1\textwidth}
    \setlength{\arrayrulewidth}{0.25mm}
    \setlength{\tabcolsep}{4pt}
    \renewcommand{\arraystretch}{1.6}
    \begin{tabular}{|ccc||ccc||ccc||c||c|}
        \hline
        \multicolumn{3}{|c||}{\textbf{Users}} & \multicolumn{3}{c||}{\textbf{Beams on Detectors}} & \multicolumn{3}{c||}{\textbf{Detector click}} & \multicolumn{1}{c||}{\textbf{Detection}} & \multicolumn{1}{c|}{\textbf{Required}} \\ \hhline{|-|-|-|-|-|-|-|-|-|}
        
        \multicolumn{1}{|c|}{$\textbf{A}$} & \multicolumn{1}{c|}{$\textbf{B}_\textbf{1}$} & $\textbf{B}_\textbf{2}$ & \multicolumn{1}{c|}{$\hat{a}_0^{\dagger}$} & \multicolumn{1}{c|}{$\hat{a}_1^{\dagger}$} & \multicolumn{1}{c||}{$\hat{a}_2^{\dagger}$} & \multicolumn{1}{p{0.028\textwidth}|}{\textbf{D0}} & \multicolumn{1}{p{0.028\textwidth}|}{\textbf{D1}} & \multicolumn{1}{p{0.028\textwidth}||}{\textbf{D2}} & \multicolumn{1}{c||}{\textbf{Type}} & \multicolumn{1}{c|}{\textbf{Actions}} \\ \hhline{|=|=|=#=|=|=#=|=|=#=#=|}
        
        \multicolumn{1}{|c|}{\textbf{Y}} & \multicolumn{1}{c|}{\textbf{X}} & \textbf{Y} & \multicolumn{1}{c|}{} & \multicolumn{1}{c|}{} & \multicolumn{1}{c||}{} & \multicolumn{1}{c|}{} & \multicolumn{1}{c|}{} & \multicolumn{1}{c||}{} & \multicolumn{1}{c||}{} & \multicolumn{1}{c|}{} \\ \hhline{|=|=|=#=|=|=#=|=|=#=#=|}
        
        \multicolumn{1}{|c|}{$\frac{\pi}{2}$} & \multicolumn{1}{c|}{0 or $\pi$} & $\frac{\pi}{2}$ & \multicolumn{1}{c|}{\multirow{2}{*}{$1.30\sqrt{\mu}$}} & \multicolumn{1}{c|}{\multirow{2}{*}{$0.39\sqrt{\mu}$}} & \multicolumn{1}{c||}{\multirow{2}{*}{$1.11\sqrt{\mu}$}} & \multicolumn{1}{c|}{\multirow{2}{*}{$\wedge$}} & \multicolumn{1}{c|}{\multirow{2}{*}{}} & \multicolumn{1}{c||}{\multirow{2}{*}{$\wedge$}} & \multicolumn{1}{c||}{\multirow{2}{*}{Type 2}} & \multicolumn{1}{c|}{\multirow{2}{*}{-}}\\ \hhline{|-|-|-|}
        
        \multicolumn{1}{|c|}{$\frac{3\pi}{2}$} & \multicolumn{1}{c|}{0 or $\pi$} & $\frac{3\pi}{2}$ & \multicolumn{1}{c|}{} & \multicolumn{1}{c|}{} & \multicolumn{1}{c||}{} & \multicolumn{1}{c|}{} & \multicolumn{1}{c|}{} & \multicolumn{1}{c||}{} & \multicolumn{1}{c||}{} & \multicolumn{1}{c|}{}\\ \hhline{|=|=|=#=|=|=#=|=|=#=#=|}
        
        \multicolumn{1}{|c|}{$\frac{\pi}{2}$} & \multicolumn{1}{c|}{0 or $\pi$} & $\frac{3\pi}{2}$ & \multicolumn{1}{c|}{\multirow{2}{*}{$0.55\sqrt{\mu}$}} & \multicolumn{1}{c|}{\multirow{2}{*}{$1.36\sqrt{\mu}$}} & \multicolumn{1}{c||}{\multirow{2}{*}{$0.92\sqrt{\mu}$}} & \multicolumn{1}{c|}{\multirow{2}{*}{}} & \multicolumn{1}{c|}{\multirow{2}{*}{$\wedge$}} & \multicolumn{1}{c||}{\multirow{2}{*}{}} & \multicolumn{1}{c||}{\multirow{2}{*}{Type 3}} & \multicolumn{1}{c|}{\multirow{2}{*}{B2 flips}}\\ \hhline{|-|-|-|}
        
        \multicolumn{1}{|c|}{$\frac{3\pi}{2}$} & \multicolumn{1}{c|}{0 or $\pi$} & $\frac{\pi}{2}$ & \multicolumn{1}{c|}{} & \multicolumn{1}{c|}{} & \multicolumn{1}{c||}{} & \multicolumn{1}{c|}{} & \multicolumn{1}{c|}{} & \multicolumn{1}{c||}{} & \multicolumn{1}{c||}{} & \multicolumn{1}{c|}{}\\ \hhline{|-|-|-|-|-|-|-|-|-|-|-|}
    \end{tabular}
\end{subtable}

\bigskip
\begin{subtable}{1\textwidth}
    \setlength{\arrayrulewidth}{0.25mm}
    \setlength{\tabcolsep}{3.8pt}
    \renewcommand{\arraystretch}{1.6}
    \begin{tabular}{|ccc||ccc||ccc||c||c|}
        \hline
        \multicolumn{3}{|c||}{\textbf{Users}} & \multicolumn{3}{c||}{\textbf{Beams on Detectors}} & \multicolumn{3}{c||}{\textbf{Detector click}} & \multicolumn{1}{c||}{\textbf{Detection}} & \multicolumn{1}{c|}{\textbf{Required}} \\ \hhline{|-|-|-|-|-|-|-|-|-|}
        
        \multicolumn{1}{|c|}{$\textbf{A}$} & \multicolumn{1}{c|}{$\textbf{B}_\textbf{1}$} & $\textbf{B}_\textbf{2}$ & \multicolumn{1}{c|}{$\hat{a}_0^{\dagger}$} & \multicolumn{1}{c|}{$\hat{a}_1^{\dagger}$} & \multicolumn{1}{c||}{$\hat{a}_2^{\dagger}$} & \multicolumn{1}{p{0.028\textwidth}|}{\textbf{D0}} & \multicolumn{1}{p{0.028\textwidth}|}{\textbf{D1}} & \multicolumn{1}{p{0.028\textwidth}||}{\textbf{D2}} & \multicolumn{1}{c||}{\textbf{Type}} & \multicolumn{1}{c|}{\textbf{Actions}} \\ \hhline{|=|=|=#=|=|=#=|=|=#=#=|}

        \multicolumn{1}{|c|}{\textbf{X}} & \multicolumn{1}{c|}{\textbf{Y}} & \textbf{X} & \multicolumn{1}{c|}{} & \multicolumn{1}{c|}{} & \multicolumn{1}{c||}{} & \multicolumn{1}{c|}{} & \multicolumn{1}{c|}{} & \multicolumn{1}{c||}{} & \multicolumn{1}{c||}{} & \multicolumn{1}{c|}{} \\ \hhline{|=|=|=#=|=|=#=|=|=#=#=|}
        
        \multicolumn{1}{|c|}{0} & \multicolumn{1}{c|}{$\frac{\pi}{2}$ or $\frac{3\pi}{2}$} & 0 & \multicolumn{1}{c|}{\multirow{2}{*}{$1.30\sqrt{\mu}$}} & \multicolumn{1}{c|}{\multirow{2}{*}{$0.39\sqrt{\mu}$}} & \multicolumn{1}{c||}{\multirow{2}{*}{$1.11\sqrt{\mu}$}} & \multicolumn{1}{c|}{\multirow{2}{*}{$\wedge$}} & \multicolumn{1}{c|}{\multirow{2}{*}{}} & \multicolumn{1}{c||}{\multirow{2}{*}{$\wedge$}} & \multicolumn{1}{c||}{\multirow{2}{*}{Type 2}} & \multicolumn{1}{c|}{\multirow{2}{*}{-}}\\ \hhline{|-|-|-|}
        
        \multicolumn{1}{|c|}{$\pi$} & \multicolumn{1}{c|}{$\frac{\pi}{2}$ or $\frac{3\pi}{2}$} & $\pi$ & \multicolumn{1}{c|}{} & \multicolumn{1}{c|}{} & \multicolumn{1}{c||}{} & \multicolumn{1}{c|}{} & \multicolumn{1}{c|}{} & \multicolumn{1}{c||}{} & \multicolumn{1}{c||}{} & \multicolumn{1}{c|}{}\\ \hhline{|=|=|=#=|=|=#=|=|=#=#=|}
        
        \multicolumn{1}{|c|}{0} & \multicolumn{1}{c|}{$\frac{\pi}{2}$ or $\frac{3\pi}{2}$} & $\pi$ & \multicolumn{1}{c|}{\multirow{2}{*}{$0.55\sqrt{\mu}$}} & \multicolumn{1}{c|}{\multirow{2}{*}{$1.36\sqrt{\mu}$}} & \multicolumn{1}{c||}{\multirow{2}{*}{$0.92\sqrt{\mu}$}} & \multicolumn{1}{c|}{\multirow{2}{*}{}} & \multicolumn{1}{c|}{\multirow{2}{*}{$\wedge$}} & \multicolumn{1}{c||}{\multirow{2}{*}{}} & \multicolumn{1}{c||}{\multirow{2}{*}{Type 3}} & \multicolumn{1}{c|}{\multirow{2}{*}{B2 flips}}\\ \hhline{|-|-|-|}
        
        \multicolumn{1}{|c|}{$\pi$} & \multicolumn{1}{c|}{$\frac{\pi}{2}$ or $\frac{3\pi}{2}$} & 0 & \multicolumn{1}{c|}{} & \multicolumn{1}{c|}{} & \multicolumn{1}{c||}{} & \multicolumn{1}{c|}{} & \multicolumn{1}{c|}{} & \multicolumn{1}{c||}{} & \multicolumn{1}{c||}{} & \multicolumn{1}{c|}{}\\ \hhline{|-|-|-|-|-|-|-|-|-|-|-|}
    \end{tabular}
\end{subtable}

\bigskip
\caption{Bases mismatch: XYY and YXX  bases choices of the users A, B$_1$ and B$_2$. }
\label{tab:a_2}
\begin{subtable}{1\textwidth}
    \setlength{\arrayrulewidth}{0.25mm}
    \setlength{\tabcolsep}{4pt}
    \renewcommand{\arraystretch}{1.6}
    \begin{tabular}{|ccc||ccc||ccc||c||c|}
        \hline
        \multicolumn{3}{|c||}{\textbf{Users}} & \multicolumn{3}{c||}{\textbf{Beams on Detectors}} & \multicolumn{3}{c||}{\textbf{Detector click}} & \multicolumn{1}{c||}{\textbf{Detection}} & \multicolumn{1}{c|}{\textbf{Required}} \\ \hhline{|-|-|-|-|-|-|-|-|-|}
        
        \multicolumn{1}{|c|}{$\textbf{A}$} & \multicolumn{1}{c|}{$\textbf{B}_\textbf{1}$} & $\textbf{B}_\textbf{2}$ & \multicolumn{1}{c|}{$\hat{a}_0^{\dagger}$} & \multicolumn{1}{c|}{$\hat{a}_1^{\dagger}$} & \multicolumn{1}{c||}{$\hat{a}_2^{\dagger}$} & \multicolumn{1}{p{0.028\textwidth}|}{\textbf{D0}} & \multicolumn{1}{p{0.028\textwidth}|}{\textbf{D1}} & \multicolumn{1}{p{0.028\textwidth}||}{\textbf{D2}} & \multicolumn{1}{c||}{\textbf{Type}} & \multicolumn{1}{c|}{\textbf{Actions}} \\ \hhline{|=|=|=#=|=|=#=|=|=#=#=|}
        
        \multicolumn{1}{|c|}{\textbf{X}} & \multicolumn{1}{c|}{\textbf{Y}} & \textbf{Y} & \multicolumn{1}{c|}{} & \multicolumn{1}{c|}{} & \multicolumn{1}{c||}{} & \multicolumn{1}{c|}{} & \multicolumn{1}{c|}{} & \multicolumn{1}{c||}{} & \multicolumn{1}{c||}{} & \multicolumn{1}{c|}{} \\ \hhline{|=|=|=#=|=|=#=|=|=#=#=|}
        
        \multicolumn{1}{|c|}{0 or $\pi$} & \multicolumn{1}{c|}{$\frac{\pi}{2}$} & $\frac{\pi}{2}$ & \multicolumn{1}{c|}{\multirow{2}{*}{$0.55\sqrt{\mu}$}} & \multicolumn{1}{c|}{\multirow{2}{*}{$0.92\sqrt{\mu}$}} & \multicolumn{1}{c||}{\multirow{2}{*}{$1.36\sqrt{\mu}$}} & \multicolumn{1}{c|}{\multirow{2}{*}{}} & \multicolumn{1}{c|}{\multirow{2}{*}{}} & \multicolumn{1}{c||}{\multirow{2}{*}{$\wedge$}} & \multicolumn{1}{c||}{\multirow{2}{*}{Type 4}} & \multicolumn{1}{c|}{\multirow{2}{*}{-}}\\ \hhline{|-|-|-|}
        
        \multicolumn{1}{|c|}{0 or $\pi$} & \multicolumn{1}{c|}{$\frac{3\pi}{2}$} & $\frac{3\pi}{2}$ & \multicolumn{1}{c|}{} & \multicolumn{1}{c|}{} & \multicolumn{1}{c||}{} & \multicolumn{1}{c|}{} & \multicolumn{1}{c|}{} & \multicolumn{1}{c||}{} & \multicolumn{1}{c||}{} & \multicolumn{1}{c|}{}\\ \hhline{|=|=|=#=|=|=#=|=|=#=#=|}
        
        \multicolumn{1}{|c|}{0 or $\pi$} & \multicolumn{1}{c|}{$\frac{\pi}{2}$} & $\frac{3\pi}{2}$ & \multicolumn{1}{c|}{\multirow{2}{*}{$1.3\sqrt{\mu}$}} & \multicolumn{1}{c|}{\multirow{2}{*}{$1.11\sqrt{\mu}$}} & \multicolumn{1}{c||}{\multirow{2}{*}{$0.39\sqrt{\mu}$}} & \multicolumn{1}{c|}{\multirow{2}{*}{$\wedge$}} & \multicolumn{1}{c|}{\multirow{2}{*}{$\wedge$}} & \multicolumn{1}{c||}{\multirow{2}{*}{}} & \multicolumn{1}{c||}{\multirow{2}{*}{Type 5}} & \multicolumn{1}{c|}{\multirow{2}{*}{B2 flips}}\\ \hhline{|-|-|-|}
        
        \multicolumn{1}{|c|}{0 or $\pi$} & \multicolumn{1}{c|}{$\frac{3\pi}{2}$} & $\frac{\pi}{2}$ & \multicolumn{1}{c|}{} & \multicolumn{1}{c|}{} & \multicolumn{1}{c||}{} & \multicolumn{1}{c|}{} & \multicolumn{1}{c|}{} & \multicolumn{1}{c||}{} & \multicolumn{1}{c||}{} & \multicolumn{1}{c|}{}\\ \hhline{|-|-|-|-|-|-|-|-|-|-|-|}
    \end{tabular}
\end{subtable}

\bigskip
\begin{subtable}{1\textwidth}
    \setlength{\arrayrulewidth}{0.25mm}
    \setlength{\tabcolsep}{3.55pt}
    \renewcommand{\arraystretch}{1.6}
    \begin{tabular}{|ccc||ccc||ccc||c||c|}
        \hline
        \multicolumn{3}{|c||}{\textbf{Users}} & \multicolumn{3}{c||}{\textbf{Beams on Detectors}} & \multicolumn{3}{c||}{\textbf{Detector click}} & \multicolumn{1}{c||}{\textbf{Detection}} & \multicolumn{1}{c|}{\textbf{Required}} \\ \hhline{|-|-|-|-|-|-|-|-|-|}
        
        \multicolumn{1}{|c|}{$\textbf{A}$} & \multicolumn{1}{c|}{$\textbf{B}_\textbf{1}$} & $\textbf{B}_\textbf{2}$ & \multicolumn{1}{c|}{$\hat{a}_0^{\dagger}$} & \multicolumn{1}{c|}{$\hat{a}_1^{\dagger}$} & \multicolumn{1}{c||}{$\hat{a}_2^{\dagger}$} & \multicolumn{1}{p{0.028\textwidth}|}{\textbf{D0}} & \multicolumn{1}{p{0.028\textwidth}|}{\textbf{D1}} & \multicolumn{1}{p{0.028\textwidth}||}{\textbf{D2}} & \multicolumn{1}{c||}{\textbf{Type}} & \multicolumn{1}{c|}{\textbf{Actions}} \\ \hhline{|=|=|=#=|=|=#=|=|=#=#=|}

        \multicolumn{1}{|c|}{\textbf{Y}} & \multicolumn{1}{c|}{\textbf{X}} & \textbf{X} & \multicolumn{1}{c|}{} & \multicolumn{1}{c|}{} & \multicolumn{1}{c||}{} & \multicolumn{1}{c|}{} & \multicolumn{1}{c|}{} & \multicolumn{1}{c||}{} & \multicolumn{1}{c||}{} & \multicolumn{1}{c|}{} \\ \hhline{|=|=|=#=|=|=#=|=|=#=#=|}
        
        \multicolumn{1}{|c|}{$\frac{\pi}{2}$ or $\frac{3\pi}{2}$} & \multicolumn{1}{c|}{0} & 0 & \multicolumn{1}{c|}{\multirow{2}{*}{$0.55\sqrt{\mu}$}} & \multicolumn{1}{c|}{\multirow{2}{*}{$0.92\sqrt{\mu}$}} & \multicolumn{1}{c||}{\multirow{2}{*}{$1.36\sqrt{\mu}$}} & \multicolumn{1}{c|}{\multirow{2}{*}{}} & \multicolumn{1}{c|}{\multirow{2}{*}{}} & \multicolumn{1}{c||}{\multirow{2}{*}{$\wedge$}} & \multicolumn{1}{c||}{\multirow{2}{*}{Type 4}} & \multicolumn{1}{c|}{\multirow{2}{*}{-}}\\ \hhline{|-|-|-|}
        
        \multicolumn{1}{|c|}{$\frac{\pi}{2}$ or $\frac{3\pi}{2}$} & \multicolumn{1}{c|}{$\pi$} & $\pi$ & \multicolumn{1}{c|}{} & \multicolumn{1}{c|}{} & \multicolumn{1}{c||}{} & \multicolumn{1}{c|}{} & \multicolumn{1}{c|}{} & \multicolumn{1}{c||}{} & \multicolumn{1}{c||}{} & \multicolumn{1}{c|}{}\\ \hhline{|=|=|=#=|=|=#=|=|=#=#=|}
        
        \multicolumn{1}{|c|}{$\frac{\pi}{2}$ or $\frac{3\pi}{2}$} & \multicolumn{1}{c|}{0} & $\pi$ & \multicolumn{1}{c|}{\multirow{2}{*}{$1.3\sqrt{\mu}$}} & \multicolumn{1}{c|}{\multirow{2}{*}{$1.11\sqrt{\mu}$}} & \multicolumn{1}{c||}{\multirow{2}{*}{$0.39\sqrt{\mu}$}} & \multicolumn{1}{c|}{\multirow{2}{*}{$\wedge$}} & \multicolumn{1}{c|}{\multirow{2}{*}{$\wedge$}} & \multicolumn{1}{c||}{\multirow{2}{*}{}} & \multicolumn{1}{c||}{\multirow{2}{*}{Type 5}} & \multicolumn{1}{c|}{\multirow{2}{*}{B2 flips}}\\ \hhline{|-|-|-|}
        
        \multicolumn{1}{|c|}{$\frac{\pi}{2}$ or $\frac{3\pi}{2}$} & \multicolumn{1}{c|}{$\pi$} & 0 & \multicolumn{1}{c|}{} & \multicolumn{1}{c|}{} & \multicolumn{1}{c||}{} & \multicolumn{1}{c|}{} & \multicolumn{1}{c|}{} & \multicolumn{1}{c||}{} & \multicolumn{1}{c||}{} & \multicolumn{1}{c|}{}\\ \hhline{|-|-|-|-|-|-|-|-|-|-|-|}
    \end{tabular}
\end{subtable}
\end{table*}
\end{appendix}

\section*{Acknowledgements}
This work was supported by European Union’s Horizon
Europe research and innovation program under grant agreement No.101092766 (ALLEGRO Project) and  Hellas QCI project co-funded by the European Union under the Digital Europe Programme  grant agreement  No.101091504.

\bibliography{IEEE_bibliography}

\end{document}